\documentclass[12pt]{iopart}
\usepackage{epsfig}
\usepackage{citesort}
\newcommand{\gtrsim}{\,\rlap{\lower4pt\hbox{$\mathchar\sim$}}
\raise1.8pt\hbox{$>$}\,}
\newcommand{\lesssim}{\,\rlap{\lower4pt\hbox{$\mathchar\sim$}}
\raise1.8pt\hbox{$<$}\,}

\begin{document}
\hbox to\hsize{\hfill MPP-2007-54}

\title{\ Collective neutrino flavor transitions in supernovae  
            \\ \ and the role of trajectory averaging }
   
\author{Gianluigi Fogli$^{1,2}$, 
		Eligio Lisi$^2$, 
		Antonio Marrone$^{1,2}$,
 		Alessandro Mirizzi$^{1,2,3}$ }
\address{$^1$~Dipartimento Interateneo di Fisica ``Michelangelo Merlin'',
		Via Orabona 4, 70126 Bari, Italy} 
\address{$^2$~Istituto Nazionale di Fisica Nucleare, Sezione di Bari,
         Via Orabona 4, 70126 Bari, Italy}
\address{$^3$~Max-Planck-Institut f\"ur Physik
(Werner-Heisenberg-Institut), F\"ohringer Ring 6, 80805 M\"unchen,
Germany} 
 

\begin{abstract}
Non-linear effects on supernova neutrino oscillations, associated with 
neutrino-neutrino interactions, are known to induce collective flavor 
transformations near the supernova core for $\theta_{13} \neq 0$. 
In scenarios with very shallow electron density profiles, 
these transformations have been shown to
couple with ordinary matter effects, jointly producing
spectral distortions both in normal and inverted hierarchy.
In this work we consider a complementary scenario, characterized
by higher electron density, as indicated by 
shock-wave simulations during a few seconds after bounce.
In this case, early
collective flavor transitions are decoupled
from later, ordinary matter effects. Moreover, such transitions
become more amenable to both numerical computations and analytical 
interpretations in inverted hierarchy, while they basically vanish 
in normal hierarchy. We numerically evolve the neutrino density matrix 
in the region relevant for self-interaction effects, using thermal spectra and
a representative value $\sin^2\theta_{13}=10^{-4}$.
In the approximation of averaged intersection angle
between neutrino trajectories, our simulations neatly show the
collective phenomena of synchronization, bipolar oscillations, and spectral
split, with analytically understandable features, as
recently discussed in the literature. In the more realistic 
(but computationally demanding) case 
of non-averaged neutrino trajectories, our simulations do not 
show new significant qualitative features, apart from the
smearing of ``fine structures'' such as bipolar nutations. 
Our results seem to
suggest that, at least for non-shallow matter density profiles, 
averaging over neutrino trajectories plays a minor role 
in the final outcome.
In this case, the swap of $\nu_e$ and $\nu_{\mu,\tau}$ spectra
above a critical energy 
may represent an unmistakable signature of the
inverted neutrino hierarchy, 
especially for $\theta_{13}$ small enough
to render further (ordinary or even turbulent) 
matter effects irrelevant.
\end{abstract}

\pacs{14.60.Pq, 13.15.+g, 97.60.Bw}

\maketitle

\section{Introduction} 

Core-collapse supernovae (SN) provide us with an interesting 
laboratory for studying both neutrino properties and their interplay with
astrophysical processes (see \cite{Raffelt:2007nv,Cavanna:2003fx} for  recent reviews). Mikheev-Smirnov-Wolfenstein (MSW)
effects induced on neutrinos by background matter \cite{Matt}
have been widely studied as a tool to probe both neutrino masses and mixings and the SN dynamics. Recent examples include the characterization,
at the level of observable SN neutrino event spectra, of shock-wave evolution effects after bounce~\cite{Schi,Taka,Luna,Foglish,Tomas,FogliMega,Barger:2005it,%
Choubey:2006aq,Dasgupta:2005wn,Kneller:2007kg} and of it possible ``erasing'' by 
stochastic matter fluctuations induced by turbulence~\cite{Fogli:2006xy,Friedland:2006ta,Choubey:2007ga}.

Ordinary MSW effects (and their ``stochastic'' smearing, if any) 
typically occur when $\omega \sim \lambda$, where  
\begin{equation}
\omega=\frac{\Delta m^2}{2E} \label{omega}
\end{equation}
is the vacuum oscillation frequency (in terms of the neutrino energy
$E$ and
squared mass difference $\Delta m^2$)%
\footnote{In this paper we neglect $\delta m^2=m^2_2-m^2_1\ll \Delta m^2$, 
where  $\Delta m^2=|m^2_3-m^2_{1,2}|$. The only relevant mixing angle is then $\theta_{13}$, governing the oscillation amplitude
in the channels $\nu_e\to\nu_x$ and 
$\overline\nu_e\to\overline\nu_x$ ($x=\mu$ or $\tau$).}
 while
\begin{equation}
\lambda(r) = \sqrt{2}\, G_F\, N_{e^-} (r) \label{lambda}
\end{equation}
is the $\nu_e$-$\nu_x$ interaction energy difference in matter,
$N_{e^-}(r)$ being the net electron number density at the point $r$. 
For typical
shock-wave density profiles as used, e.g., in \cite{Foglish,Tomas}, the condition
$\omega \sim \lambda$ occurs after a few hundred (or even 
a few thousand) kilometers, during the first few seconds after core bounce.
Significantly shallower electron density profiles (as those adopted in \cite{Duan:2006an}
in the context of models with successful $r$-process nucleosynthesis) can 
instead trigger MSW effects much earlier, around $O(100)$~km.

Besides electrons, neutrinos can also be a nontrivial 
background to themselves when their density is large enough 
\cite{Fuller,Pantaleone:1992eq}.
Self-interaction effects,
being inherently non-linear, are very different
(and much less intuitive) than ordinary MSW effect, and can lead to 
collective flavor transition 
phenomena in which neutrinos (or antineutrinos) of any energy behave
similarly~\cite{Samuel:1993uw,Kostelecky:1993yt,Kostelecky:1993dm,%
Kostelecky:1994ys,%
Kostelecky:1995dt,Kostelecky:1995xc,Samuel:1996ri,%
Kostelecky:1996bs,Pantaleone:1998xi,%
Abazajian:2002qx,Dolgov:2002ab,Wong:2002fa,%
Pastor:2001iu,Sawyer:2005jk,%
Sigl:2007yz}.
The interest of such effects for the neutrino flavor evolution
in the dense SN core has long been recognized~\cite{Pantaleone:1994ns,%
Qian:1994wh,Sigl:1994hc,Pastor:2002we,Balantekin:2004ug}. 
 Roughly speaking, 
significant self-interaction phenomena are expected when 
$\mu(r) \gtrsim \omega$, where
\begin{equation}
\mu(r) = \sqrt{2}\, G_F\, [N(r)+\overline N(r)] \label{mu}
\end{equation}
$N(r)$ and $\overline N(r)$ being the total effective
neutrino ($\nu_e+\nu_x$) and
antineutrino ($\overline\nu_e+\overline\nu_x$) number density,
respectively (to be precisely defined later). 

In the most general case, systems with dense matter and dense neutrino gases
are thus governed by (at least) three characteristic frequencies:
$\omega$ (spread over $\sim 2$ orders of magnitude for typical
energy spectra); $\lambda$ (roughly decreasing as the third power of the distance); and $\mu$
(decreasing as the fourth power of the distance \cite{Duan:2006an}). Neutrino flavor evolution
becomes then a complicated multi-scale dynamical problem with a rich
phenomenology, which may involve both collective and MSW effects, the latter
being often assumed (at least in older literature) to lead to flavor transitions. 
In the SN neutrino context, this ``old'' paradigm
(focussing on MSW effects) has dramatically changed 
after the emergence of dominant collective phenomena (such as the so-called
``bipolar oscillations'') studied in detailed, large-scale computer simulations
\cite{Duan:2006an,Duan:2006jv} as well as in simplified but
analytical models
\cite{Fuller:2005ae,Duan:2005cp,Hannestad:2006nj,Duan:2007mv,RaffSmirn,%
Duan:2007fw,%
Duan:2007bt}.

The interaction strength between two (anti)neutrinos is modulated by a factor
$(1-\cos\vartheta_{ij})$, where $\vartheta_{ij}$ is the angle between their
intersecting trajectories.
If one ignores the spread of $\vartheta_{ij}$, and averages it out along a ``representative''  
radial trajectory (single-angle approximation), 
and if one also assumes that self-interaction effects do not
interfere with the ordinary MSW ones, then the following picture appears
to emerge in supernovae
from analytical considerations. Nothing relevant occurs for
normal hierarchy ($m_3>m_{1,2}$), while, for inverted mass hierarchy ($m_3<m_{1,2}$),
any value of $\theta_{13}\neq 0$
(no matter 
how small~\cite{Duan:2005cp,Hannestad:2006nj})
can trigger 
collective pair-conversions of the kind 
$\nu_e\overline\nu_e\to \nu_x\overline \nu_x$ 
within the first $O(100)$ km.
Then, as recently emphasized 
in~\cite{RaffSmirn,Duan:2007fw}, the end of collective effects 
is marked by a spectral pair-conversion
which is complete for $\overline\nu$'s, while for $\nu$'s it occurs
only above a characteristic energy set by lepton
number conservation. 
Such spectral ``split'' (or ``stepwise swap'' of flavors)
would then represent an unmistakable signature of self-interaction
effects~\cite{Duan:2006an,Duan:2006jv,Duan:2007mv,RaffSmirn,Duan:2007fw,Duan:2007bt,RafLast}. 
Its robustness needs, however, to be  better investigated in 
increasingly refined SN models including, e.g., 
variable neutrino crossing angles $\vartheta_{ij}$ (multi-angle simulations), 
which might induce kinematical decoherence effects, 
(``depolarization'' and ``smearing of oscillations''), which are 
neither obvious nor entirely clear in the few numerical 
\cite{Duan:2006an,Esteban-Pretel:2007ec} and analytical \cite{Hannestad:2006nj,Sigl:2007yz}
studies performed so far. 
In our scenario, it turns out that main results are
rather robust when passing from single- to multi-angle
simulations. 

The purpose of this paper is to explore such ``self-interaction dominated'' scenario  in
a realistic case characterized by: (1) an appropriate matter profile, where collective effects
fully develop before MSW effects (if any); (2)
continuous, thermal energy spectra with significant neutrino-antineutrino (and neutrino flavor)
asymmetry, and (3) numerical simulations in single- and multi-angle cases. In this sense, our work is complementary
to the simulations in \cite{Duan:2006an}, where the shallower adopted
matter profile allows MSW effects to occur
within (not beyond) the range of collective transitions---which leads
to a richer phenomenology, but also to more
difficult analyses and much greater
numerical challenges. 
In the terminology of \cite{Duan:2005cp}, we study the scenario with ``thick'' rather
than ``thin'' envelope. 

The plan of our work is as follows.
In Section~2 we describe our supernova reference model.
In Section~3 we set the notation and write
the neutrino evolution equations in single-angle approximation.
In Section~4 and 5 we discuss single-angle analytical and numerical solutions,
respectively. In 
Section~5 we tackle multi-angle simulations,
and show that the main final effect (the spectral split) is robust.
Conclusions and perspectives are presented in Section~6.
Technical aspects are discussed in the Appendix.

\section{Reference Supernova Model}

Our  reference model is characterized by the following choices for the
initial neutrino energy spectra, the
geometry and intensity of neutrino emission, and the
matter profile, which govern the distribution of the three basic ``frequencies''
$\omega$, $\mu$, and $\lambda$,
respectively. 

We shall use
such choices in the usual, simplified context of pure two-family ($\nu_e,\,\nu_x$) evolution
\cite{Duan:2005cp,Hannestad:2006nj}, where
$\nu_x$ represents a single active flavor. It is worth noticing that, for $\lambda\neq 0$, 
this case does not exactly represent the two-family limit (for $\delta m^2=0)$ of the 
general three-family case, contrary to the familiar
MSW effects in SNe. In fact, in the presence of self-interactions,
the flavor evolution depends on the absolute effective neutrino densities and, thus, also on
the total number of neutrino families (either two or three) assumed to share the total luminosity. 
The full $3\nu$ case (and its proper $2\nu$ limit) 
will be studied elsewhere.   

\subsection{Initial Neutrino Energy Spectra}

We assume normalized thermal spectra with different temperatures
$T=1/\beta$ for $\nu_e$, $\overline\nu_e$, and $\nu_x$ (the latter 
having the same properties as $\overline\nu_x$). More precisely, the initial 
energy spectra $\phi^i(E)$ are of the form 
\begin{equation}
\phi^i(E) = \frac{2\beta}{3\zeta_3} 
\;\frac{(\beta E)\,^2}{e^{\beta E } +1} \,\ ,\label{phi}
\end{equation}
where $\zeta_3\simeq 1.202$. The average values of
$E$ and $E^{-1}$ are then:
\begin{equation}
\langle E^{\pm 1}\rangle = \int dE\, \phi(E)\, 
E^{\pm 1} = c_{\pm}\,T^{\pm 1}\ = c_{\pm}\,\beta^{\mp 1} ,
\label{averages}
\end{equation}
where $c_+=7\pi^4/180\zeta_3\simeq 3.151$ and $c_-=\pi^2/18\zeta_3\simeq 0.4561$.
In numerical calculations we choose  $\langle E_e \rangle = 10$~MeV, 
$\langle \overline E_e \rangle = 15$~MeV, 
and $\langle E_x \rangle = \langle \overline E_x \rangle = 24$~MeV
for $\nu_e$, $\overline\nu_e$, $\nu_x$ and $\overline \nu_x$,
respectively, corresponding to 
\begin{equation}
\beta_e=0.315\,,\ \overline\beta_e=0.210\,,\ \beta_x=\overline\beta_x=0.131\ (\mathrm{MeV}^{-1})\ .
\end{equation}

\subsection{Emission Geometry and Intensity}

We adopt the``bulb model''  emission  described in~\cite{Duan:2006an},
to which the reader is referred for further details. Neutrinos
are assumed to be half-isotropically emitted above the neutrino-sphere,
i.e., all the outward moving modes are equally occupied as expected from
a blackbody emission. The
neutrino number flux $j_\nu$ per unit energy (in any direction) is
then given by \cite{Duan:2006an}
\begin{equation}
{j_\nu}(E)  = \frac {F_\nu}{2\pi}\, \frac{\phi^i(E)}
{\langle E \rangle}\label{jnu}\ ,
\end{equation}
where
\begin{equation}
F_{\nu} = \frac{L_\nu}{2\pi R_\nu^2}\ ,
\end{equation}
$R_\nu$ being the neutrino-sphere radius, while 
$L_\nu$ is the total emission power for a given neutrino
species. In numerical calculations, we assume reference values
$R_\nu=10$~km and
$L_\nu=10^{51}$ erg/s for each species $\nu=\nu_e$, $\overline\nu_e$,
 $\nu_x$, $\overline\nu_x$. 

\begin{figure}[t]
\centering
\vspace*{-4mm}
\hspace*{14mm}
\epsfig{figure=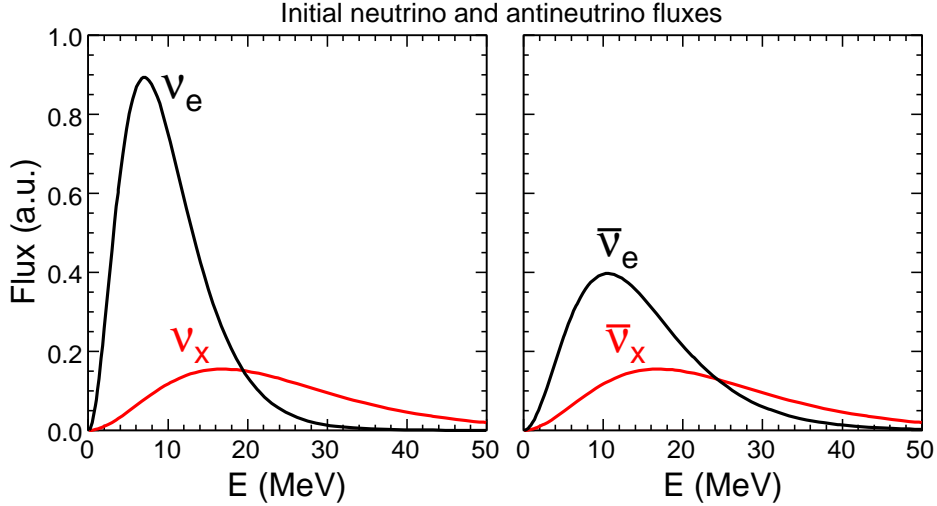,width =1.00\columnwidth}
\vspace*{-10mm}
\caption{Initial fluxes (at $r=10$~km, in arbitrary units) 
for different neutrino species as a function of energy. The fluxes are 
all proportional to $\phi^i(E)/\langle E\rangle$.
\label{fig1}}
\end{figure}

Figure~\ref{fig1} shows the initial neutrino number fluxes per unit energy
in arbitrary units (all fluxes being proportional to $\phi^i(E)/\langle E\rangle$ 
through the same normalization constant). Notice the significant difference (asymmetry) between neutrinos
and antineutrinos, and between different neutrino flavors.
However, the $\nu_e$ and $\nu_x$ fluxes happen to coincide at an energy
$E_\mathrm{eq}\simeq 19$~MeV, while for the $\overline\nu_e$ and 
$\overline\nu_x$ fluxes the equality occurs at $\overline E_\mathrm{eq}\simeq 
24$~MeV. Flavor transformations of any kind
are not operative for neutrinos at $E=E_{\mathrm{eq}}$,
and for antineutrinos at $E=\overline{ E}_{\mathrm{eq}}$.

The spherical symmetry of emission reduces to a cylindrical symmetry
along the radial line-of-sight (polar axis). 
At any radius $r>R_\nu$ along the polar axis, neutrinos will 
arrive with different momenta ${\bf p}$ characterized by 
$|{\bf p}|=E$, incident polar angle $\vartheta$, and
azimuthal angle $\varphi$. In the calculation of self-interaction
effects, the effective differential neutrino number density 
$dn_{\bf p}$ with momentum between ${\bf p}$ and
${\bf p} + d{\bf p}$ is then \cite{Duan:2006an} 
\begin{equation}
dn_{\bf p} = j_{\nu} (E) d\Omega = j_{\nu} (E)\,d \varphi\, d\cos \vartheta \ ,
\end{equation}
within the cone of sight of the neutrino-sphere,  with $\vartheta \in [0,\vartheta_{\max}]$, being 
\begin{equation}
\vartheta_{\max}=\arcsin (R_\nu/r) \ .\label{varthetamax}
\end{equation}

In general, angular coordinates are important, since the 
interaction strength between two neutrinos of momenta ${\bf p}$ and ${\bf q}$ 
depends on their relative angle $\vartheta_{\bf pq}$ through
the factor $(1-\cos\vartheta_{\bf pq})$. Calculations embedding
the full angular coordinates are dubbed ``multi-angle.''
The often used ``single-angle'' 
approximation consists in averaging the angular
factor along the 
polar axis, which is assumed to encode the same flavor history 
of any other neutrino direction. In this case, 
the effective neutrino number 
density $n$ per unit volume and energy turns out to be  \cite{Duan:2006an}
\begin{equation}
n(r,E)=2\pi\, D(r)\,j_\nu(E) = F_\nu \,D(r)\,\frac{\phi^i(E)}{\langle E\rangle}\label{n}
\end{equation}
for the various neutrino species ($n=n_e,\,\overline n_e,\, n_x$ or $\overline n_x$),
where the (species-independent) geometrical function $D(r)$ is given by
\begin{equation}
D(r)=\frac{1}{2}\left[1-\sqrt{1-\left(
\frac{R_\nu}{r}\right)^2} \right]^2\ ,
\end{equation}
decreasing as $r^{-4}$ for large $r$. The $\nu$-$\overline\nu$
asymmetry of the model implies $n_e\neq \overline n_e$.

It is useful to integrate
the effective densities per unit energy and volume ($n_e$, $\overline n_e$,
and $n_x=\overline n_x$) to obtain the effective number densities of
$\nu_e$, $\overline\nu_e$ and $\nu_x$ ($\overline\nu_x)$,
\begin{eqnarray}
N_e &=&\int dE\, n_e = F_\nu\frac{D(r)}{\langle E_e \rangle}
=\frac{F_\nu}{c_+}\,D(r)\,\beta_e\ ,\\
\overline N_e &=&\int dE\, \overline n_e=F_\nu
\frac{D(r)}{\langle \overline E_e \rangle} =\frac{F_\nu}{c_+}\,D(r)\,\overline\beta_e\ ,\\
N_x &=& \int dE\, n_x =F_\nu\frac{D(r)}{\langle E_x \rangle} =\frac{F_\nu}{c_+}\,D(r)\,\beta_x\ = \overline N_x
\end{eqnarray}
as well as the total effective number densities of neutrinos and antineutrinos,
\begin{eqnarray}
N&=&N_e+N_x\ ,\\
\overline N&=&\overline N_e+\overline N_x\ ,
\end{eqnarray}
from which one can finally derive the 
parameter $\mu=\sqrt{2}\, G_F\, (N+\overline N)$, which governs the
neutrino self-interaction strength.

\begin{figure}[t]
\centering
\vspace*{-10mm}
\hspace*{18mm}
\epsfig{figure=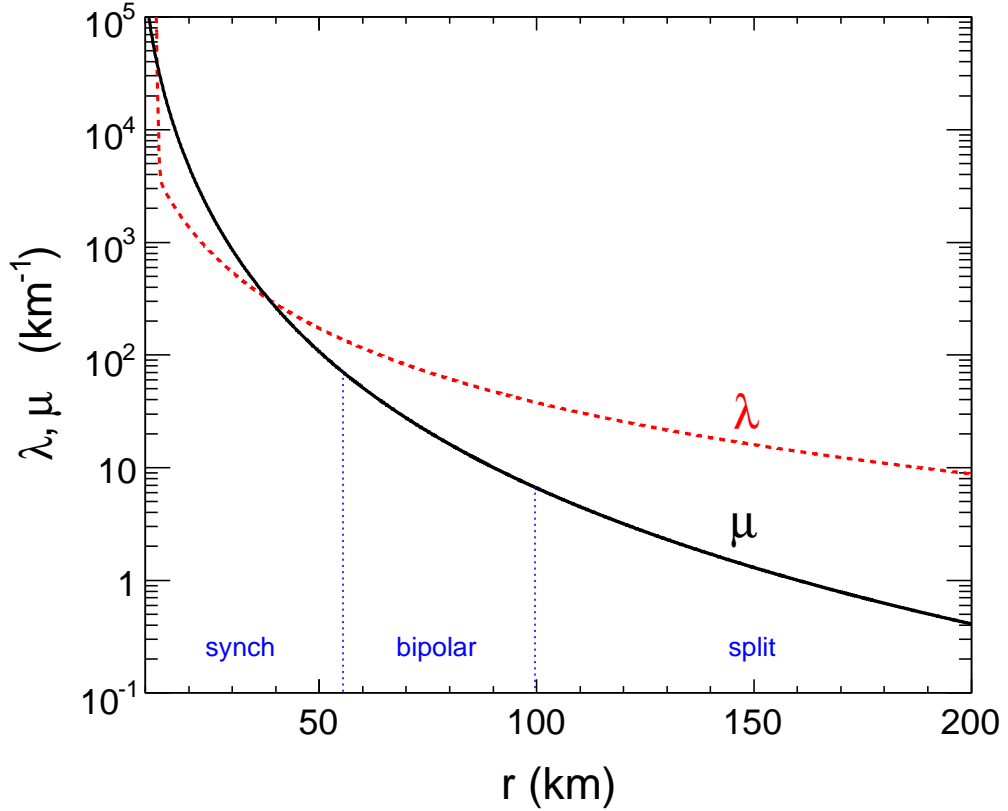,width =1.0\columnwidth}
\vspace*{-5mm}
\caption{Radial profiles of the neutrino self-interaction parameter
$\mu(r)=\sqrt{2}\,G_F\,(N+\overline N)$ and of the matter-interaction
parameter $\lambda(r)=\sqrt{2}\,G_F\,N_{e^-}$ adopted in this work,
in the range $r\in[10,\,200]$~km. 
Shown are also the approximative
ranges where self-interaction effects are expected to
produce mainly synchronization, bipolar oscillations
and spectral split (see the text for details).
\label{fig2}}
\end{figure}

Figure~\ref{fig2} shows the function $\mu(r)$ in our reference SN model (together
with the ordinary MSW strength $\lambda(r)$ defined below)
in the range $r\in[10,\,200]$~km relevant for self-interaction effects.
Also shown are the approximate ranges where these effects induce
synchronization, bipolar oscillations and spectral split,
as discussed later in Sec.~3.

\subsection{Matter Profile} 

Shock waves propagating during the first few second after bounce can
produce very interesting MSW effects when $\lambda\sim \omega$, provided
that $\sin^2\theta_{13}$ is not too small \cite{Schi,Taka,Luna,Foglish,Tomas,FogliMega,Barger:2005it,Choubey:2006aq}. 
For the typical
(time-dependent) matter profiles used in these studies,
such effects usually emerge in the first $\sim 10$~s after a few hundred kilometers 
from the SN core, in which case collective neutrino self-interaction 
effects are already completely developed, as we shall see later. 
In this case, the specific choice of the neutrino potential profile
$\lambda(r)=\sqrt{2}G_fN_{e^-}$  turns out to have only 
a minor impact on the evolution of self-interaction effects, even if $\lambda(r) \gg \mu (r)$.  
For the sake of definiteness, we single out one 
of the time-dependent profiles studied
in~\cite{Foglish} (the one at post-bounce time $t=5$~s),
and steepen it in a few km range above the neutrino sphere 
as suggested in~\cite{Duan:2006an,Qian:1993dg}.%
\footnote{The local steepening does not change any feature
of the collective neutrino flavor transformations,
but it turns out to help the start-up of our numerical evolution routines 
in the first few km.}
The resulting profile $\lambda(r)$ adopted in this work
is shown in Figure~\ref{fig2}.

\section{Single-Angle Approximation: Analytical aspects}

An ensemble of relativistic neutrinos and antineutrinos 
coming in  $F$ flavors can be described by a set of dimensionless
$F \times F$ density matrices $\rho_{{\bf p}}$ and $\bar{\rho}_{{\bf p}}$,
one for each momentum
mode. The most general Liouville
evolution equations for 
$\rho_{{\bf p}}$ have been worked out in~\cite{Sigl:1992fn}. 
In this Section we specialize and discuss such equations in Bloch form
for $F=2$ and single-angle approximation. Although we
mostly rely on the formalism 
introduced in \cite{Hannestad:2006nj} and on the results presented in 
\cite{Duan:2005cp,Duan:2006an,Hannestad:2006nj,Duan:2007mv,RaffSmirn}, 
we think it useful---for the sake of clarity and self-consistency---to give a systematic overview of the 
asymmetric ($n_e\neq \overline n_e$) 
and continuous-energy case, also because our choice of thermal spectra
allows some useful analytical estimates in terms
of temperature parameters $\beta=1/T$.

\subsection{Bloch Vector Notation}

We consider two-family mixing between $\nu_e$ and $\nu_x$ 
($\overline \nu_e$ and $\overline\nu_x$), driven by mass-mixing
parameters $(\Delta m^2,\sin^2\theta_{13})$. In this case, one can
switch from normal to inverted mass hierarchy
 in two ways: (1) mixing is kept unaltered while 
$+\Delta m^2 \to  - \Delta m^2$; 
or (2) $\Delta m^2$ is kept positive, while
the mixing angle  is swapped, $\sin\theta_{13}
\leftrightarrow \cos\theta_{13}$ (which implies $+\cos2\theta_{13}\to
-\cos2\theta_{13}$, with unaltered $\sin2\theta_{13}$). Hereafter
we adopt the latter choice, as advocated in \cite{Hannestad:2006nj}. In numerical calculations, we assume reference values
\begin{equation}
\Delta m^2=2\times 10^{-3} \mathrm{\ eV}^2\,,\ \ \sin^2\theta_{13}=10^{-4}\ .
\end{equation}

In the single-angle approximation, the only kinematical parameter is
$E=|{\bf p}|$.
For each energy $E$ and radius $r$, the 
neutrino density matrix $\rho$ in flavor basis can be written 
in terms of polarization (Bloch) vector ${\bf P}=(P_x,\,P_y,\,P_z)^T=P_x{\bf x}+P_y{\bf y}+P_z{\bf z}$, being
${\bf x},{\bf y}$ and ${\bf z}$ three orthogonal unit vectors,
 and of the vector of Pauli matrices 
$\mbox{\boldmath$\sigma$}=(\sigma_x,\,\sigma_y,\,\sigma_z)^T$, 
\begin{equation}\label{eq:h2}
\rho =  \left( \begin{array}{cc}
\rho_{ee} & \rho_{ex} \\ \rho_{xe} & \rho_{xx} \end{array} \right)
=  \left( \begin{array}{cc}
|\nu_e|^2 & \nu_e \nu_x^\ast  \\ \nu_e^\ast \nu_x & |\nu_x|^2 \end{array} \right)
=\frac{n}{2} \left( {\bf 1} + {\bf P} \cdot \mbox{\boldmath$\sigma$} \right) \,\ ,
\end{equation}
where $\bf 1$ is the unit matrix, $n= \Tr(\rho)=n_e+n_x$ represents the occupation number of neutrinos 
(per unit volume) at energy $E$, and
the densities are defined through Eq.~(\ref{n}) in our model.
Analogous definitions in terms of $\overline {\bf P}$ hold for antineutrinos.
The initial  conditions in flavor basis
$\rho^i={\rm diag}(n_e,\,n_x)$ and ${\overline \rho}^i=
{\rm diag}(\overline n_e,\,\overline n_x)$ correspond to 
\begin{eqnarray}
{\bf P}^i&=&P^i_z\,{\bf z}=\frac{n_e-n_x}{n}\,{\bf z}\ ,\\
{\overline{\bf P}}^i&=&\overline{P}^i_z\,{\bf z}=\frac{\overline n_e-\overline n_x}{\overline n}\,{\bf z}\ .
\end{eqnarray}
The final ($f$) survival probabilities are then given by
\begin{eqnarray}
P_{ee}&=&P(\nu_e^i\to \nu_e^f)=\frac{1}{2}\left(1+
\frac{P_z^f}{P_z^i}\right)\ ,\\ 
\overline{P}_{ee}&=&P(\overline\nu_e^i\to \overline\nu_e^f)=\frac{1}{2}\left(1+
\frac{\overline{P}_z^f}{\overline{P}_z^i}\right)\ .
\end{eqnarray}

It is useful to introduce the integral polarization vectors
of neutrinos and antineutrinos,
\begin{eqnarray}
{\bf J}&=&\frac{1}{N+\overline N}\int dE\,n\,{\bf P}\ , \\
{\bf \overline J}&=&\frac{1}{N+\overline N}\int dE\,\overline n\,{\bf\overline P}\ , 
\end{eqnarray}
as well as their sum ${\bf S}$ and difference ${\bf D}$,
\begin{eqnarray}
{\bf S} &=& {\bf J}+{\bf\overline J}\ ,\\
{\bf D} &=& {\bf J}-{\bf\overline J}\ .  
\end{eqnarray}
The initial conditions imply that
\begin{eqnarray}
{\bf J}^i &=& \frac{N_e-N_x}{N+\overline N}\,{\bf z}=
\frac{\beta_e-\beta_x}{\beta_e+\overline\beta_e+2\beta_x}\,{\bf z}\ ,\\
{\bf \overline J}^i &=& \frac{\overline N_e-\overline N_x}{N+\overline N}\,{\bf z}
=\frac{\overline\beta_e-\beta_x}{\beta_e+\overline\beta_e+2\beta_x}\,{\bf z}\ ,\\
{\bf S}^i &=& \frac{N_e+\overline N_e-2N_x}{N+\overline N}\,{\bf z}
=\frac{\beta_e+\overline\beta_e-2\beta_x}{\beta_e+\overline\beta_e+2\beta_x}\,{\bf z}\ ,\\
{\bf D}^i &=& \frac{N_e-\overline N_e}{N+\overline N}\,{\bf z}
=\frac{\beta_e-\overline\beta_e}{\beta_e+\overline\beta_e+2\beta_x}\,{\bf z}\ .
\end{eqnarray}

Another auxiliary (unit) vector is the ``magnetic field,'' 
\begin{equation}
{\bf B}=\sin 2\theta_{13}\,{\bf x}\mp \cos2\theta_{13}\,{\bf z}
\end{equation}
where the upper (lower) sign refers to  normal
(inverted) hierarchy. For small $\theta_{13}$, one can take
${\bf B}\simeq \mp{\bf z}$, unless the relevant dynamics is very
close to the $z$ axis.

\subsection{Equations Of Motion And Removal of Matter Effects}

The Bloch equations of motion (EOM)
for each polarization vector ${\bf P}$ 
 read, in single-angle approximation,
\begin{eqnarray}
\dot \mathbf{P}&=&\left(+\omega {\bf B}+\lambda {\bf z}+\mu {\bf D}\right)\times {\bf P}\ ,\label{Bloch1}\\
\dot \mathbf{\overline P}&=&\left(-\omega {\bf B}+\lambda {\bf z}+\mu {\bf D}\right)\times {\bf\overline P}\ ,\label{Bloch2}
\end{eqnarray}
where the three terms in brackets
embed vacuum, matter, and self-interaction effects. 
In particular, the third term couples all modes ${\bf P}$ and $\overline {\bf P}$,
and is responsible for collective effects.
The equations conserve each $|{\bf P}|$ and thus unitarity. It is understood
that ${\bf P}={\bf P}(E,r)$, $\lambda=\lambda(r)$, and $\mu=\mu(r)$, with $r=t$ and $d_t=d_r$.
If the continuous parameter $E$ is discretized through a set
of $N_E$ points $\{E_h\}_{h=1,\dots,N_E}$,  then a set
of $6\times N_E$ coupled, first-order differential equations in $t$
is obtained. The equations are ``stiff,'' namely, 
their solutions generally involve a fast-changing combination of
multi-frequency oscillations,
due to the ``precession''
of ${\bf P}$ and ${\bf \overline P}$ around the three terms in brackets.
It is thus amazing that, through appropriate approximations, the resulting dynamics turns out to be understandable in terms of simple phenomena, as discussed below.

In a frame rotating with angular velocity $\lambda {\bf z}$ 
\cite{Duan:2005cp},
the time derivative acquires an extra operator $-\lambda {\bf z} \times$,
which cancels the matter term. Moreover, the $z$-component of ${\bf P}$
(and of any other vector) are unchanged in such co-rotating frame,
and thus the survival probability $P_{ee}$ is also unchanged. Only
the transverse $(x,y)$ components of any vector get
mixed, e.g., the ``magnetic field'' becomes
${\bf B} = (\sin 2\theta_{13} \cos(-\lambda t),\, 
\sin 2 \theta_{13} \sin(-\lambda t),\, \mp \cos 2 \theta_{13})^T$. Again,
one can still take
${\bf B}\simeq {\bf \mp z}$, unless 
the relevant dynamics does not occur too close to ${\bf B}$ or ${\bf z}$
(a case which will be discussed separately). Matter $(\lambda)$ effects thus
``disappear'' in such co-rotating frame, 
as pointed out in~\cite{Duan:2005cp},
leaving the other terms
in the EOM formally unchanged:
\begin{eqnarray}
\dot \mathbf{P}&=&\left(+\omega {\bf B}+\mu {\bf D}\right)\times {\bf P}\ ,
\label{EOMP}\\
\dot \mathbf{\overline P}&=&\left(-\omega {\bf B}+\mu {\bf D}\right)\times {\bf\overline P}\ . \label{EOMPbar}
\end{eqnarray}

The corresponding EOM for integral quantities are:
\begin{eqnarray}
\dot {\bf J} &=& +{\bf B}\times {\bf W}+\mu {\bf D}\times {\bf J} \ ,
\label{EOMJ}\\
\dot {\bf \overline J} &=& -{\bf B}\times {\bf \overline W}+\mu {\bf D}\times {\bf \overline J} \ ,\label{EOMJbar}\\
\dot {\bf S} &=& {\bf B}\times ({\bf W-\overline W})+\mu {\bf D}\times {\bf S}\ ,\label{EOMS}\\
\dot {\bf D} &=& {\bf B}\times ({\bf W+\overline W})\ ,\label{EOMD}
\end{eqnarray}
where 
\begin{eqnarray}
{\bf W} &=& \frac{1}{N+\overline N}\int dE \,\omega\, n\, {\bf P}\ ,
\label{W}\\
{\bf \overline W} &=& \frac{1}{N+\overline N}\int dE \,\omega\, \overline n\, 
{\bf \overline P}\ ,\label{Wbar}
\end{eqnarray}
with initial conditions implying
\begin{eqnarray}
{\bf W}^i &=&\frac{1}{N+\overline N}\int dE 
\,\omega\, (n_e-n_x)\, {\bf z} = \frac{\Delta m^2 \,c_-}{2}\,
\frac{\beta_e^2-\beta_x^2}{\beta_e+\overline \beta_e+2\beta_x}
\,{\bf z}\ ,\\
{\bf \overline W}^i &=& \frac{1}{N+\overline N}\int dE 
\,\omega\, (\overline n_e-\overline n_x)\, {\bf z}
=\frac{\Delta m^2 \,c_-}{2}\,
\frac{\overline \beta_e^2-\beta_x^2}{\beta_e+\overline \beta_e+2\beta_x}
\,{\bf z} \ .
\end{eqnarray}

\subsection{Conservation Laws and Spectral Split}

The equation for $\dot {\bf D}$ implies conservation of the scalar 
\begin{equation}
{\bf D}\cdot{\bf B}=\mathrm{const}={\bf D}^i\cdot{\bf B}\simeq
\mp{\bf D}^i\cdot{\bf z}=\mp\frac{N_e-\overline N_e}{N+\overline N}\ ,
\end{equation}
corresponding to the conservation of the (electron) lepton number,
and implying pair conversions of the kind $\nu_e\overline\nu_e\to \nu_x\overline \nu_x$ \cite{Hannestad:2006nj}.

In the special case of (nearly) constant neutrino density
($\dot \mu\simeq 0$), another conserved scalar
is the average energy per neutrino pair, given by
\begin{equation}
{\cal E}= {\bf B}\cdot({\bf W+\overline W})+\frac{1}{2}\mu {\bf D}^2= {\cal V}+{\cal T}\ ,
\end{equation}
where the first (second) term acts as a sort of potential (kinetic) energy.

When $r$ is sufficiently large to make self-interaction effects vanish
($\mu\ll\omega$), the kinetic term ${\cal T}$ also vanishes, and 
the ``ground state'' for ${\cal E}$ would corresponds
to the minimization of the potential $\cal V$, i.e., to $\bf W$  and $\bf \overline W$
aligned as much as possible with $-\bf B$ (in any hierarchy),
provided that lepton number is also conserved.

Since the vectors ${\bf W}$ and ${\bf \overline W}$ always start aligned
to ${\bf z}$, this implies that, in normal
hierarchy (${\bf z}\simeq -{\bf B}$) they  end up in the same position,
trivially conserving lepton number (i.e., the system starts---and remains---close
to the minimum of the potential energy).
Conversely, in inverted hierarchy (${\bf z}\simeq +{\bf B}$) 
the vectors ${\bf W}$ and ${\bf \overline W}$ start antialigned
with $-{\bf B}$ (maximum of the potential), and for large $r$ they  
should reverse their direction in order to approach the potential minimum.
Reversal cannot be complete, however, since it would maximally violate
(invert) lepton number.
Consistent minimization of the potential
can be achieved by
complete reversal of the smallest
vector $\bf \overline W$, and by partial
reversal of  the
largest vector ${\bf W}$ (note that the excess of neutrinos over antineutrinos leads to
$|{\bf W}|>|{\bf \overline W}|$).

More precisely, only a fraction ${\bf W}_{>}$ of ${}\bf W$, the fraction 
above a certain critical (split) energy $E_c$, is reversed, while
the complementary fraction, ${\bf W}_{<}={\bf W}-{\bf W}_{>}$, remains 
unaltered \cite{RaffSmirn,Duan:2007mv,Duan:2007fw,Duan:2007bt,RafLast,Esteban-Pretel:2007ec}. 
In such final state, the critical energy is fixed by lepton number conservation,
 i.e., by $D_z^i=D_z^f$ with
\begin{equation}
(N+\overline N)D_z^f=\int_0^{E_c}dE(n_e-n_x)
-\int_{E_c}^\infty dE(n_e-n_x)
+\int_0^{\infty }dE(\overline n_e-\overline n_x)\ ,
\end{equation}
the last two terms having the opposite overall sign in the initial
state $(N+\overline N)D_z^i$. Then
one gets an implicit equation for $E_c$,
\begin{equation}
\int_{E_c}^\infty dE(n_e-n_x)=\int_0^\infty dE(\overline n_e-\overline n_x)\ ,
\end{equation}
which, in our specific SN model, is solved for $E_c\simeq 7$~MeV. When all collective effects are terminated ($\mu\ll\omega$),
one expects a nearly complete inversion of the
polarization vectors for $E>E_c$, 
and thus a ``stepwise swap'' between the $\nu_e$ and $\nu_x$
energy spectra.

Of course, such reasoning is heuristic, and does not prove that the dynamics
allows the system to reach the peculiar final state discussed above. 
We refer the reader to~\cite{Duan:2007mv,RaffSmirn,Duan:2007fw,Duan:2007bt,RafLast} for explicit  
solutions constructed in adiabatic approximation
(slowly decreasing $\mu$), which indeed lead to spectral split 
under rather broad assumptions. 
On the other hand, such adiabatic solutions
average out the interesting  transient phenomenon of bipolar oscillations \cite{Duan:2005cp,Hannestad:2006nj},
which we discuss below.

\subsection{Alignment Approximation and Gyroscopic Pendulum}

For $\mu |{\bf D}|\gg \omega$, Eqs.~(\ref{EOMP})--(\ref{EOMJbar}) reduce to
the same form $\dot {\bf V}\simeq \mu{\bf D}\times {\bf V}$, and thus
all polarization vectors $\bf V=P,\,\overline P,\, J,$ and $\overline{\bf J}$ 
have the same dynamics (in particular, they
remain closely aligned to each other, and to the $z$-axis, 
as they are at the start). As $\mu$ decreases, the $\pm\omega$ terms in the EOM
start to be non-negligible, and neutrino and antineutrino polarization  
vectors develop different precession histories. 
As far as $\Delta \omega/\mu$ remains small,
where $\Delta\omega$ is the typical energy spread, the individual ${\bf P}$'s
stick to the global 
vector ${\bf J}$ (i.e., their components parallel to ${\bf J}$
typically dominate over the transverse ones), and analogously 
the $\overline{\bf P}$'s stick to $\overline{\bf J}$, with
${\bf J}$ and $\overline{\bf J}$ gradually separating from each
other.%
\footnote{This approximation may become ill-defined in the 
symmetric case $n_e=\overline n_e$ (not our case), 
where the smallness of $|\bf D|$ makes 
the condition $\mu |{\bf D}|\gg \omega$ critical
and the $({\bf J},\overline{\bf J})$ separation unclear. See,
e.g., the remarks in Sec.~VI~A of \protect\cite{Hannestad:2006nj}.}

Within such ``alignment approximation,'' also
${\bf W}$ ($\overline{\bf W}$) is nearly parallel to 
 ${\bf J}$ ($\overline{\bf J}$),
\begin{eqnarray}
{\bf W} &\simeq & w \,{\bf J}\ ,\\
\overline {\bf W} &\simeq & \overline w\, \overline{\bf J}\ ,
\end{eqnarray}
and Eqs.~(\ref{EOMJ})-(\ref{EOMJbar}) become 
\begin{eqnarray}
\dot {\bf J} &=& [(\omega_\mathrm{dif}+\omega_\mathrm{ave}){\bf B}+\mu {\bf D}]\times {\bf J} \ ,
\label{EOMJ1}\\
\dot {\bf \overline J} &=& [(\omega_\mathrm{dif}-\omega_\mathrm{ave}){\bf B}+\mu {\bf D}]\times {\bf \overline J} \ ,\label{EOMJbar1}
\end{eqnarray}
where we have defined
\begin{eqnarray}
\omega_\mathrm{ave}&=&(w+\overline w)/2\ ,\\
\omega_\mathrm{dif}&=&(w-\overline w)/2 \ .
\end{eqnarray}

Equations~(\ref{EOMJ1})-(\ref{EOMJbar1}) imply conservation  
of the vector moduli $J=|{\bf J}|$ and $\overline J=|\overline{\bf J}|$, 
as well as 
 of $W=|\mathbf{W}|$ and $\overline W=|\overline\mathbf{W}|$. 
The frequencies $w$ and $\overline w$ can be evaluated, e.g., in
the initial state, providing
\begin{eqnarray}
\omega_\mathrm{ave}&=&\frac{1}{2}\left(\frac{W^i_z}{J^i_z}+\frac{\overline W^i_z}{\overline J^i_z} \right)
=\frac{\int dE\, \omega(n_e-n_x)}{2(N_e-N_x)}+
\frac{\int dE\, \omega(\overline n_e-\overline n_x)}{2(\overline N_e-
\overline N_x)}\nonumber\\
&=&\frac{\Delta m^2 c_-}{4}(\beta_e+\overline\beta_e+2\beta_x)\ .
\end{eqnarray}
In our specific SN model, for $\Delta m^2=2\times 10^{-3}$ eV$^2$ it is
\begin{equation}
\omega_\mathrm{ave}\simeq 0.9 \ \mathrm{km}^{-1}\ .
\end{equation}
By going in a co-rotating frame with frequency 
$\omega_\mathrm{dif}{\bf B}\simeq \mp \omega_\mathrm{dif}{\bf z}$, the 
terms $\omega_\mathrm{dif}{\bf B} \times$ in Eqs.~(\ref{EOMJ1})-(\ref{EOMJbar1})
 are rotated away, with no other formal change in the
EOM for $\bf J$ and $\overline{\bf J}$.

The $\omega_\mathrm{dif}{\bf B} \times$ terms disappear also
from the EOM of $\bf S$ and $\bf D$. By defining 
${\bf Q}={\bf S}-(\omega_\mathrm{ave}/\mu){\bf B}$ \cite{Hannestad:2006nj},
{\em and\/} assuming $\dot \mu\simeq 0$ (adiabatic variations of $\mu$),
Eqs.~(\ref{EOMS})-(\ref{EOMD}) can be written as
\begin{eqnarray}
\dot{\bf Q}=\mu {\bf D}\times {\bf Q} \ , \label{EOMQD}\\
\dot{\bf D}=\omega_\mathrm{ave} {\bf B}\times {\bf Q}\label{EOMDQ}\ ,
\end{eqnarray}
showing that the $(\nu,\overline\nu)$ ensamble is characterized 
by a single, collective 
frequency $\omega_\mathrm{ave}$, despite the existence of a
continuous energy spectra. Notice that
$Q=|{\bf Q}|$ is conserved, as well as ${\bf D}\cdot {\bf B}$ and
${\bf D}\cdot {\bf Q}$  \cite{Hannestad:2006nj}. 

It has been realized that Eqs.~(\ref{EOMQD})-(\ref{EOMDQ})
describe a gyroscopic pendulum in flavor space \cite{Hannestad:2006nj,Duan:2007mv}.
In particular, by making the identifications
\begin{eqnarray}
{\bf Q}/Q &\equiv& {\bf r} \ (\mathrm{unit\ length\ vector})\ ,\\
{\bf D} &\equiv& {\bf L} \ (\mathrm{total\ angular\ momentum})\ ,\\
{\mu ^{-1}} &\equiv&  m \ (\mathrm{mass})\ ,\\
{\bf D}\cdot {\bf Q}/Q &\equiv& \sigma \ (\mathrm{spin})\ ,\\
\omega_\mathrm{ave}\, \mu\, Q\, {\bf B} &\equiv& -{\bf g} \ (\mathrm{gravity\ field})\ ,
\end{eqnarray} 
one can write Eqs.~(\ref{EOMQD})-(\ref{EOMDQ}) in the form (after
right-multiplying Eq.~(\protect\ref{EOMQD}) by ${\bf r}\times$)
\begin{eqnarray}
{\bf L}&=&m{\bf r}\times\dot{\bf r}+\sigma{\bf r}\ ,\\
\dot{\bf L}&=& m{\bf r}\times{\bf g}\ ,
\end{eqnarray}
which are the equations of motion of a 
spherical pendulum of unit length ($|{\bf r}|=1$), subject
to a constant gravity field ${\bf g}$, and characterized by
a point-like bob of mass $m$ which spins around the pendulum axis $\bf r$
with constant (inner) angular momentum $\sigma$. The most general 
evolution of this system is a combination of two periodic motions of the bob, one
around the vertical ${\bf g}$ axis (precession) and the other
along it (nutation) \cite{Hannestad:2006nj,Duan:2007mv}. An explicit solution 
of the EOM can be constructed
in terms of quadratures involving
the three integrals of motion \cite{Goldstein,Landau}, namely, the 
spin $\sigma ={\bf L}\cdot {\bf r}$, the vertical component of
the angular momentum
${\bf L}\cdot {\bf g}/|{\bf g}|$, and the energy ${\cal E}$ which now
equals
\begin{equation}
{\cal E}=-m\, {\bf g}\cdot{\bf r}+\left(\frac{m}{2}\,{\dot{\bf r}}^2
+\frac{\sigma^2}{2m}\right)\ ,
\end{equation}
where we have kept the spin term in the (bracketed) kinetic energy.
The explicit solution, however, involves elliptic integrals and
is not particularly transparent. Here it is sufficient to recall  the
following global features of the pendulum motion.

In normal hierarchy, the pendulum starts close to the stable, downward 
position (the misalignment being of $O(\theta_{13})$), and remains
close to it as $\mu$ slowly decreases (i.e., $m$ slowly increases).
Conversely, in inverted hierarchy, the pendulum starts close to the ``unstable,''
upward position. When $\mu$ is large and thus $m$ is small, however, the 
bob spin $\sigma$ dominates (``fast rotator''), and the pendulum remains precessing in the
upward position to conserve angular momentum (``sleeping top'') \cite{Duan:2007mv,Goldstein}. 
This situation (also dubbed as ``synchronization'' \cite{Pastor:2001iu,Hannestad:2006nj}
in the SN neutrino context) is stable if the dominant (spin) kinetic
term is larger than the maximum excursion of the potential energy
\cite{Hannestad:2006nj,Duan:2007mv,Goldstein}, namely, 
\begin{equation}
\frac{\sigma^2}{2m}>2m|{\bf g}|\ .
\end{equation}
For large $\mu/\omega_\mathrm{ave}$ (and thus vertically aligned 
polarization vectors), 
this condition translates into $\mu>\mu_\mathrm{sup}$, where
\begin{eqnarray}
\mu_\mathrm{sup}\simeq \frac{4\,\omega_\mathrm{ave}\,S_z^i}{(D_z^i)^2}&=&
4\,{\omega_\mathrm{ave}}\,\frac{(N_e+\overline N_e)^2-(N_x+\overline N_x)^2}{(N_e-\overline N_e)^2}\nonumber \\
&=&
4\,{\omega_\mathrm{ave}}\,\frac{(\beta_e+\overline \beta_e)^2-
4\beta_x^2}{(\beta_e-\overline \beta_e)^2}\ .\end{eqnarray}
In our reference SN model, it is 
\begin{equation}
\mu_\mathrm{sup} \simeq  75 \,\omega_\mathrm{ave} \simeq 67 \ {\mathrm{km}}^{-1}\ .
\end{equation}

Conversely, when $\mu<\mu_\mathrm{sup}$, any initial misalignment with the
``vertical axis'' set by $\bf g$ (i.e., any $\theta_{13}\neq 0$, no matter how small) triggers
the first fall of the pendulum and its subsequent  
nutations (also dubbed as ``bipolar oscillations''
in the SN neutrino context). If $\mu$ were constant, bipolar
oscillations would be exactly periodic.
However, the decrease of $\mu$ implies an increase of
 the pendulum inertia ($m$); the pendulum never swings back to 
exactly the same uppermost position, which instead steadily decreases, 
together with the vertical amplitude of nutations
(roughly $\propto {\mu}^{1/2}$ \cite{Hannestad:2006nj,Duan:2007mv}).
Bipolar oscillations are then expected to vanish when self-interaction 
and vacuum effects are comparable, 
and the ``alignment approximation'' breaks down.
More precisely, one may expect this condition to occur when the
$\mu$ and $\omega$ terms in the EOM of $\bf J$
and $\overline{\bf J}$ are of the same size
($\mu  {\bf D}\cdot{\bf B}\sim \omega_\mathrm{ave}$), implying
$\mu\sim \mu_\mathrm{inf}$ with 
\begin{equation}
\mu_\mathrm{inf}= \frac{\omega_\mathrm{ave}}{D_z^i}=
\omega_\mathrm{ave}\frac{N+\overline N}{N_e-\overline N_e}=
\omega_\mathrm{ave}\frac{\beta_e+\overline \beta_e+2\beta_x}
{\beta_e-\overline \beta_e}
\end{equation}
In our reference SN model, it is 
\begin{equation}
\mu_\mathrm{inf} \simeq  7.5 \,\omega_\mathrm{ave} \simeq 6.7 \ {\mathrm{km}}^{-1}\ .
\end{equation}

The condition $\mu\simeq \mu_\mathrm{inf}$ roughly marks the ``end'' of
bipolar oscillations and of collective effects, but not yet of all self-interaction effects. 
In particular,
spectral split effects continue to 
build up for
$\mu\lesssim \mu_\mathrm{inf}$, 
and eventually freeze out for $\mu \ll \omega_{\rm{ave}}$. The reason is
that the neutrino spectral split requires a separation of the
vector $\bf W$ into parts
$\bf W_{<}$ and $\bf W_{>}$ oppositely evolving in the
$z$-component. As far as the alignment approximation holds
(and thus bipolar oscillations occur), this split cannot
fully develop, and should therefore be complete somewhat beyond
the bipolar range. Of course, there is no sharp boundary
between the two processes: in the range where $\mu\sim \mu_\mathrm{inf}$,
one should observe a smooth vanishing of bipolar oscillations, and a
gradual build-up of the spectral split, through
the polarization reversal of neutrinos with $E>E_c$, with an
associated non-conservation of $\bf W$ (and of $\bf J$). 
Summarizing, we expect the following sequence of dominant phenomena, where the 
radial ranges refer to our reference SN model:
\begin{eqnarray}
\mu\gtrsim\mu_\mathrm{sup} &:&
\mathrm{synchronized\ oscillations} \ (r\lesssim 55\mathrm{\ km})\ ,\\
\mu_\mathrm{inf}\lesssim\mu\lesssim\mu_\mathrm{sup} &:&
\mathrm{bipolar\ oscillations} \ (55{\mathrm{\ km}}\lesssim
r\lesssim 100\mathrm{\ km})\ ,\\
\mu\lesssim\mu_\mathrm{inf} &:&
\mathrm{spectral\ split} \ (r \gtrsim100{\mathrm{\ km}})\ .
\end{eqnarray}
Such ranges are explictly shown in Fig.~\ref{fig2}.
For numerical purposes, we shall stop our investigations to 200 km in this paper. 
We are not concerned here with subsequent (ordinary or ``stochastic'') MSW
transitions which may occur later at $r\sim O(10^3)$~km
when $\lambda\simeq \omega$.

\subsection{The Revenge of Matter Effects}

Matter $\lambda$ effects can alter the previous description in two ways:
(1) by anticipating the MSW condition $\lambda\simeq \omega_\mathrm{ave}$ 
within the first $O(200)$~km in shallow matter profiles (not our case),
thus interfering with collective effects in the same range \cite{Duan:2006an,Duan:2007fw};(2)
by altering the dynamics when the polarization vectors are very close
to ${\bf z}$, which occurs just at the transition between synchronized and
bipolar oscillations \cite{Hannestad:2006nj}. When the latter transition occurs, 
matter ($\lambda$) and small mixing $(\theta_{13})$
effects cannot be completely co-rotated away,
the ${\bf B}\simeq \mp{\bf z}$ approximation fails, and
the transverse components of $\bf B$ (oscillating with amplitude 
$\sin2\theta_{13}$ and frequency $\lambda$) must be taken into account.
In the nontrivial case of inverted hierarchy, it turns out that
their general effect is to further stabilize  the ``upward'' pendulum
position \cite{Hannestad:2006nj}, elongating the period of (at least) the first bipolar swing.
Explicit analytic estimates have been presented in \cite{Hannestad:2006nj} for the symmetric case
$n_e=\overline n_e$, by
solving the full EOM with time-varying $\bf B$ and for small deviations around
the vertical directions. If we take these estimates as a reasonable proxy also for our asymmetric case ($n_e\neq \overline n_e$), we expect that the first 
bipolar oscillation (nutation)
should be a factor of $\sim 4$ longer than the next ones (which are much less
affected by these subtle matter effect, the uppermost pendulum
position becoming increasingly tilted for decreasing $\mu$). The
onset of bipolar oscillations is then expected to be somewhat delayed
beyond $r=55$~km in our reference SN model, by a time span equivalent to 
``a few nutations.''

\section{Single-Angle Approximation: Numerical results}

The previous analytical expectations for the single-angle case
are nicely confirmed by our simulations. We numerically evolve 
Eqs.~(\ref{Bloch1})-(\ref{Bloch2}) in the range $r\in [10,\,200]$~km 
within our reference SN model, considering
only in the nontrivial case of inverted hierarchy (we have anyway checked
that no significant effect occurs in normal hierarchy).
Technical details
are discussed in the Appendix: here we focus only on the results
and their interpretation.

\begin{figure}[t]
\centering
\vspace*{-8mm}
\hspace*{15mm}
\epsfig{figure=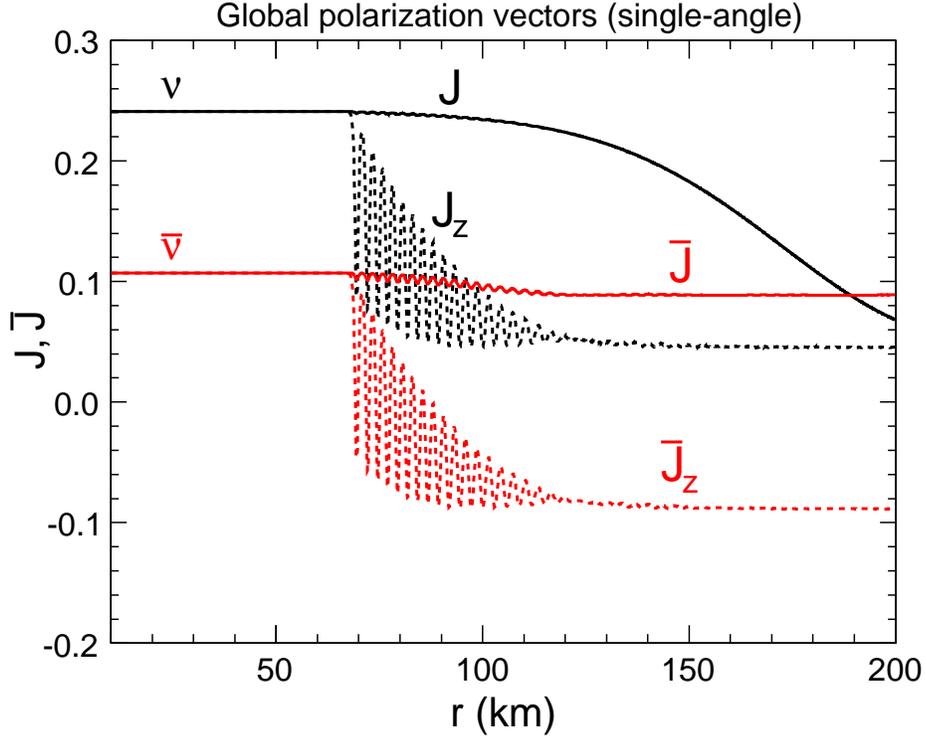,width =0.95\columnwidth}
\vspace*{-8mm}
\caption{Single-angle simulation in inverted hierarchy: Modulus and $z$-component of the global polarization vector
of neutrinos $(\bf J)$ and antineutrinos $(\overline{\bf J})$, as a function of radius. 
The difference 
$D_z=J_z-\overline{J}_z$ (not shown) remains strictly constant in $r$.
\label{fig3}}
\end{figure}

Figure~\ref{fig3} shows the radial evolution of the modulus $J$ and
$z$-component $J_z$ of the global neutrino polarization vector
$\bf J$ (and analogously for the antineutrino vector $\overline{\bf J}$). 
Their radial profile can be interpreted as follows.  
Up to 
$\sim 68$~km,
it is $J=J_z$ and $\overline J=\overline J_z$: all polarization
vectors are ``glued'' (synchronized) along the vertical axis, and the gyroscopic
flavor pendulum just spins in the upward position without falling.
At $r\sim 68$~km, the pendulum falls for the first time
and nutations appear, marking the transition from
synchronized to bipolar regime. The
transition is retarded by a few 
nutation periods (from the expected $\sim 55$~km to $\sim 68$~km) by the matter effects discussed in the previous
Section. The nutation amplitude gradually decrease
(as $\sim \mu^{1/2}$), and bipolar oscillations eventually vanish
for $r\gtrsim 100$~km, as expected. 

At the same time, the 
spectral split effect builds up. Antineutrinos tend to completely
reverse the polarization vector $(\overline{\bf J}\to -\overline{\bf J})$,
thus minimizing their ``potential energy'' (after which nothing relevant
happens to them), so that $\overline J_z\simeq-\overline J$ asymptotically.%
\footnote{There is actually a slight loss of $\overline J$ during
bipolar oscillations in Fig.~\protect\ref{fig3} (as also numerically observed in \cite{Esteban-Pretel:2007ec}) which makes
the final  $|J_z^f|\simeq 0.09$ slightly smaller than the initial $J_z^i\simeq 0.1$.}
Neutrinos also try to invert their global polarization vector (as much
as it is allowed by lepton number conservation) as soon as the alignment 
approximation breaks down ($\mu\lesssim \mu_\mathrm{inf}$) and non-conservation 
of $J$ is allowed. Indeed,
for $r\gtrsim 100$~km, $J$ decreases.
  Eventually
the situation $J\simeq J_z$  is reached (slightly beyond the $r$ range in Fig.~\ref{fig3}),
when the spectral splitting is frozen, 
corresponding to a final state with $\bf J$ aligned with $\bf +z$
 and $\overline{\bf J} $ aligned with $\bf -z$. In all the above
 processes,
lepton number is strictly conserved, 
leading to
 the constancy of 
$D_z=J_z-\overline J_z$ at any $r$. From the point of view of
observable oscillation probabilities (related to the $z$-component of
polarization vectors), the situation is basically frozen well before
$r\sim 200$~km.
 We can conclude that the behavior of $J$, 
$\overline{ J}$, $J_z$ and $\overline J_z$ is well understood,
with good agreement between analytical expectations and numerical
simulations.
The agreement can also be extended to more detailed
features of the energy spectrum, as discussed next.

\begin{figure}[t]
\centering
\vspace*{-3mm}
\hspace*{18mm}
\epsfig{figure=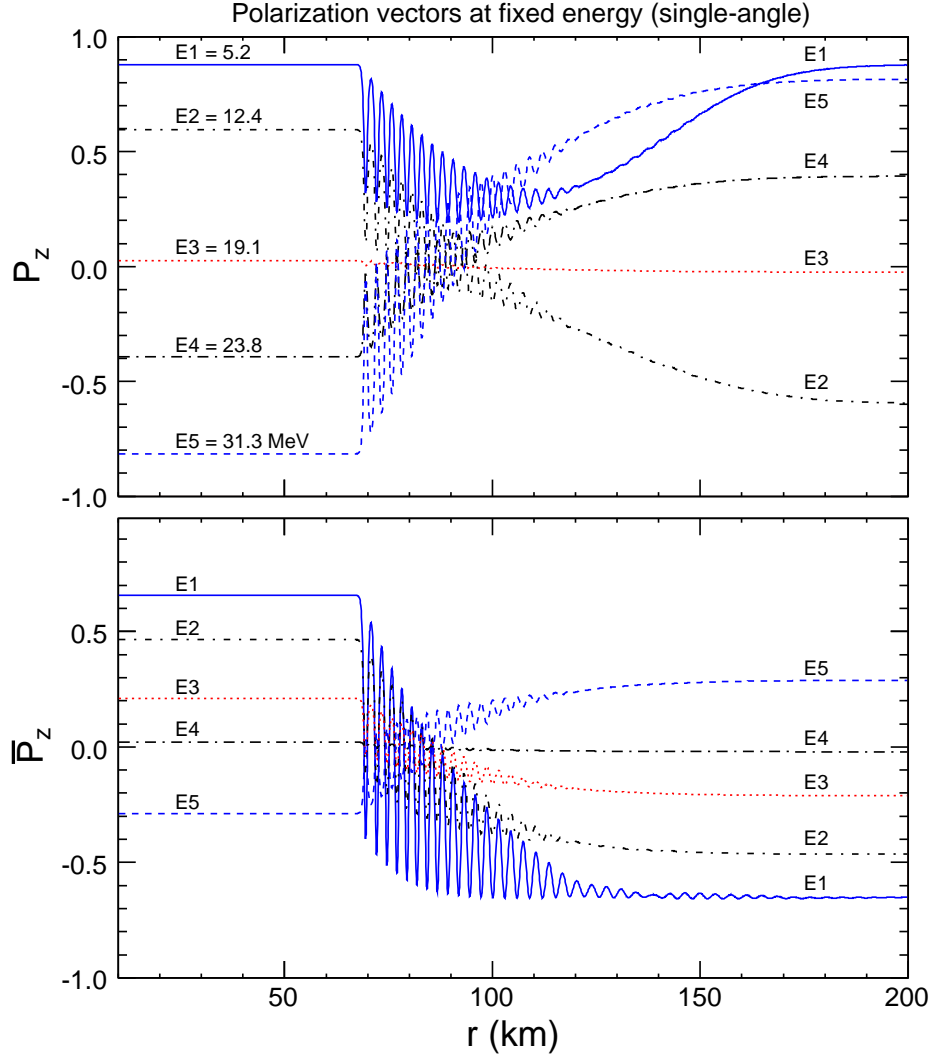,width =1.05\columnwidth}
\vspace*{-6mm}
\caption{Single-angle simulation in inverted hierarchy: $z$-component of the polarization vector
of neutrinos ($P_z$, upper panel)  and antineutrinos ($\overline P_z$, lower panel)
as a function of radius, for five representative values of the energy.
\label{fig4}}
\end{figure}

Figure~\ref{fig4} shows the behavior of the individual polarization components $ P_z$
(upper panel) and $\overline{P}_z$ (lower panel) as a function of $r$, for
five representative values of energy (in MeV): $E_1\simeq 5.2$, $E_2\simeq 12.4$,
$E_3\simeq 19.1$, $E_4\simeq 23.8$, and $E_5\simeq 31.3$ (which are a subset
of the grid sampling energies, hence their ``non-rounded'' 
values).%
\footnote{We do not show the moduli, which are always strictly conserved.}
In Fig.~\ref{fig4}, the onset of
bipolar oscillations and their nutation periods are clearly equal for both $\nu$
and $\overline\nu$ at any energy, confirming 
the appearance  of a 
self-induced collective behavior, characterized by a single
frequency parameter $\omega_\mathrm{ave}$ 
for all (anti)neutrinos, despite
the spread of vacuum oscillation frequencies $\omega$. 
However, the fate of
each $P_z$ or $\overline{P}_z$ does depend on energy. 
For neutrinos (upper panel in Fig.~\ref{fig4}), one
can clearly see the phenomenon of spectral split
around $E_c\simeq 7$ MeV:
the curve at $E_1<E_c$ ends up at the same initial value, 
while the curves at $E_2,\,E_4,\,E_5>E_c$ show the
expected inversion $P_z\to -P_z$. Only the curve at energy
$E_3$ does not change much ($P_z\simeq 0$ at any $r$),
being close to the energy $E_\mathrm{eq}$ where the $\nu_e$ and 
$\nu_x$ fluxes are equal (see Fig.~\ref{fig1}), and flavor transformations are
inoperative. 
For antineutrinos (lower panel in Fig.~\ref{fig4}), all curves show
complete polarization reversal as expected $(\overline{P}_{z}\to-\overline{P}_z)$, including the 
``trivial'' case $\overline{P}_z\simeq 0$ at
$E_4\simeq \overline E_\mathrm{eq}$, where the
$\overline\nu_e$ and $\overline \nu_x$ fluxes are equal.
We conclude that the numerical simulations confirm the
end of bipolar oscillations and the appearance of the energy split phenomenon (around 
$r\sim 100$~km) with the expected global features. There is only a minor
``unexpected'' effect (lack of $\overline{P}_z$ reversal for $E\lesssim4$~MeV, not
shown in Fig.~\ref{fig4}),
as commented below.

At $r=200$~km, self-induced flavor transformations have basically ended. The transition from
initial ($r=10$~km) to final ($r=200$~km) fluxes implies, 
in our two-family scenario 
($P_{ee}=P_{xx}=1-P_{ex}$, see also the remarks at the beginning of Sec.~2),
\begin{eqnarray}
\frac{\phi^i_e(E)}{\langle E_e\rangle} \
\longrightarrow\
\frac{\phi^i_e(E)}{\langle E_e\rangle}P_{ee}(E)+
\frac{\phi^i_x(E)}{\langle E_x\rangle}[1-P_{ee}(E)]  \ ,\\
\frac{\phi^i_x(E)}{\langle E_x\rangle} \
\longrightarrow\
\frac{\phi^i_x(E)}{\langle E_x\rangle}P_{ee}(E)+
\frac{\phi^i_e(E)}{\langle E_e\rangle}[1-P_{ee}(E)]  \ ,
\end{eqnarray}
and similarly for antineutrinos.

\begin{figure}[t]
\centering
\vspace*{-4mm}
\hspace*{14mm}
\epsfig{figure=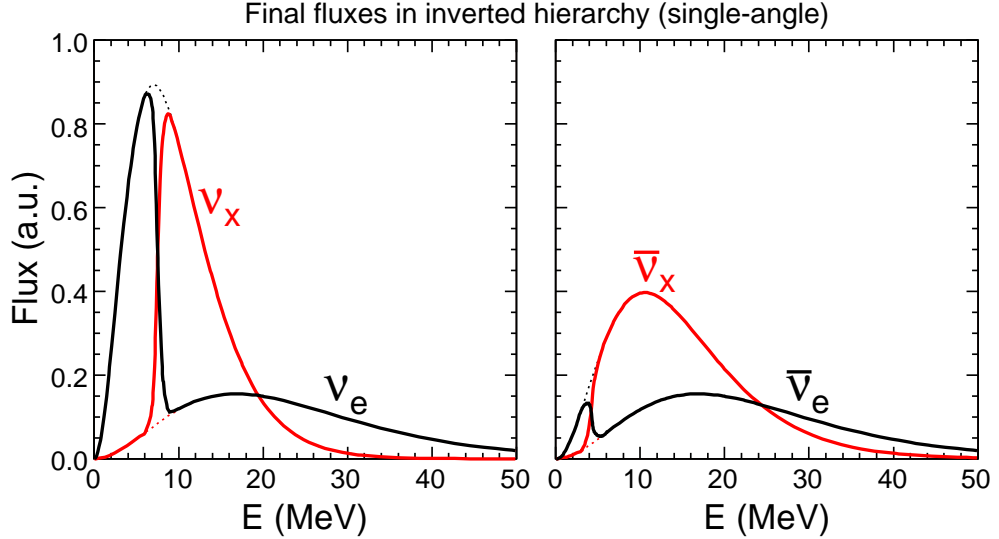,width =1.06\columnwidth}
\vspace*{-10mm}
\caption{Single-angle simulation in inverted hierarchy: 
Final fluxes (at $r=200$~km, in arbitrary units) 
for different neutrino species as a function of energy. Initial fluxes are shown
as dotted lines to guide the eye.
\label{fig5}}
\end{figure}

Figure~\ref{fig5} shows the final neutrino and antineutrino fluxes calculated in this way. 
The left panel (neutrinos) clearly shows the spectral split effect, and the 
corresponding sudden swap of $\nu_e$ and $\nu_x$ fluxes above $E_c\simeq 7$~MeV.
In the right panel of Fig.~\ref{fig5}, the final antineutrino spectra are basically
completely swapped with respect to
the initial ones (compare with Fig.~\ref{fig1}), except at very low energies, where there 
appears an ``antineutrino'' spectral split.
 We relate this phenomenon to the loss of $\overline J$ 
and of $|\overline{J}_z|$ observed and commented in Fig.~\ref{fig3}: The small
deficit $|\overline{J}^f_z|<\overline{J}^i_z$ can indeed obtained,
analogously to the neutrino case, through the
lack of $\overline{P}_z$ reversal for low-energy antineutrinos
($E<\overline E_c$ with  $\overline E_c\lesssim 4$~MeV). 
A better understanding of this minor feature and of the $|\overline{J}_z|$ loss would be desirable;
however, we anticipate that the antineutrino spectral split is largely smeared out in multi-angle simulation, contrary to 
the neutrino spectral split which appears to be a robust, observable feature.

\section{Multi-angle Simulations: Notation and Numerical results}

In the multi-angle case (applied to the neutrino bulb model),
any single polarization vector depends not only on the energy $E$ and on
the total propagation distance along a neutrino trajectory $t$,
 but also on the (incident) angle $\vartheta$ between
the neutrino trajectory and the polar axis. The vectors 
${\bf P}_\vartheta(E,t)$ and ${\overline{\bf P}}_\vartheta(E,t)$  obey then the following EOM \cite{Duan:2006an},
\begin{eqnarray}
\dot {\bf P}_\vartheta &=& \!\!\left[+\omega{\bf B}+\!\lambda{\bf z} +\!2\pi\sqrt{2}G_F\!
\int\! d c_{\vartheta'}\, dE\, (1-c_\vartheta c_{\vartheta'})(j{\bf}{\bf P}_{\vartheta'}-\overline j{\overline{\bf P}}_{\vartheta'})
\right]\!\times\!{\bf P}_\vartheta\, ,\label{bloch3}\\
\dot {\overline{\bf P}}_\vartheta &=& \!\!\left[-\omega{\bf B}+\!\lambda{\bf z} +\!2\pi\sqrt{2}G_F\!
\int\! d c_{\vartheta'}\, dE\, (1-c_\vartheta c_{\vartheta'})(j{\bf}{\bf P}_{\vartheta'}-\overline j{\overline{\bf P}}_{\vartheta'})
\right]\!\times\!{\overline{\bf P}}_\vartheta\, ,\label{bloch4}\end{eqnarray}
where $c_\vartheta=\cos\vartheta$ with $\vartheta\in[0,\vartheta_{\max}]$
and $c_{\vartheta'}=\cos\vartheta'\in[\cos\vartheta_{\max},1]$,%
\footnote{The angles $\vartheta$ and $\vartheta'$ must
lie in the cone subtending the neutrinosphere. The factor $2\pi=\int d\varphi$
comes from cylindrical symmetry within this cone.}
while
$j$ and $\overline j$ are the total neutrino and antineutrino
number densities
\begin{eqnarray}
j &=& j_e + j_x\ ,\\
\overline j &=& \overline j_e +\overline j_x\ , 
\end{eqnarray}
as defined in Sec.~2.2.

It is useful to characterize all the properties of the neutrino beam by using the
radius $r$ and   the emission angle $\vartheta_0$ at the neutrinosphere \cite{Duan:2006an}, 
by means of
\begin{eqnarray}
r\,\sin\vartheta &=& R_\nu\,\sin\vartheta_0\ \,\ , \\
t &=&  \sqrt{r^2 -R_\nu^2 \sin^2 \vartheta_0}- R_\nu \cos \vartheta_0 \,\ ,
\end{eqnarray}
the range of $\cos\vartheta_0$ being constantly $[0,1]$ at any $r$.
Along a generic trajectory at angle $\vartheta$,  $dt= dr/c_\vartheta$, and 
thus
$d_t=c_\vartheta d_r$. 
Since any couple $(t,\vartheta)$ is in one-to-one correspondence
with $(r,\vartheta_0)$, the polarization vectors can be
relabeled as ${\bf P}_{\vartheta_0}(E,r)$. 

For convenience, one may define effective densities $n$ as in the
single-angle case,
$n(r,E)=2\pi D(r) j(E)$, so that the initial conditions at $r=R_\nu$ become as usual
\begin{eqnarray}
{\bf P}_{\vartheta_0}^i &=& (j_e-j_x)/j = (n_e-n_x)/n\ ,\\
 {\overline{\bf P}}_{\vartheta_0}^i &=& (\overline j_e-\overline j_x)/\overline
 j = (\overline n_e-\overline n_x)/\overline n\ .
\end{eqnarray}
The single-angle limit is 
recovered by fixing $\vartheta=0=\vartheta_0$ and by assuming
that all polarization vectors behave as the ones at $\vartheta_0=0$,
in which case the integral $\int dc_{\vartheta'}(1-c_{\vartheta'})=D(r)$
is factorized out (this is actually the way
$D(r)$ is originally defined \cite{Duan:2006an}).
\footnote{An alternative single-angle case has been recently
 studied
in~\cite{Esteban-Pretel:2007ec}, by selecting the emission angle 
$\vartheta_0=
\pi/4$ instead of $\vartheta_0=
0 = \vartheta$.}

Equations~(\ref{bloch3})-(\ref{bloch4}) reduce then
to Eqs.~(\ref{Bloch1})-(\ref{Bloch2}) after the integrated densities 
$N$ and $\overline N$ are introduced as in Sec.~2.2.

 Angle-averaged
polarization vectors can be defined as
\begin{equation}
{\bf P}=
\frac{\int d c_{\vartheta_0}\,c_{\vartheta_0}\, {\bf P}_{\vartheta_0}}
{\int d c_{\vartheta_0}\,c_{\vartheta_0}}\label{angleave}\ ,
\end{equation}
(and similarly for $\overline{\bf P}$), where the ``extra''
cosine factor accounts for projection in radial
direction \cite{Esteban-Pretel:2007ec}. Global polarization
vectors $\bf J$, $\overline{\bf J}$ and $\bf D=J-\overline J$ 
can then be defined in the same way as in Sec.~3.2. Although their 
EOM are not as simple as in the single-angle case, it turns out
that ${\bf D}\cdot{\bf B}\simeq D_z$ is still a conserved scalar 
\cite{Hannestad:2006nj,Esteban-Pretel:2007ec}
(proof omitted). 

The angular dependence of the neutrino-neutrino interaction
strength is generally expected to introduce some ``dephasing'' 
or kinematical decoherence between different neutrino trajectories,
and to smear out the ``fine structures''  observed in single-angle
simulations. While decoherence effects are dominant
in the symmetric case $(n_e=\overline{n}_e)$ \cite{Sigl:2007yz}, they seem to
be only subdominant in asymmetric cases $(n_e\neq \overline{n}_e)$
\cite{Duan:2006an,Esteban-Pretel:2007ec}. In the latter case their quantitative description lacks, at the moment, 
of an analytical understanding, and relies mainly on 
numerical simulations~\cite{Esteban-Pretel:2007ec}.
However, one may at least expect that the appearance of the neutrino spectral split
phenomenon is not spoiled in multi-angle cases, 
 being based on the broad-brush picture of 
potential energy minimization (i.e., final (anti)alignment with $\bf B$)
constrained by lepton number conservation (i.e., constant $D_z$). 
This expectation is confirmed by the numerical simulations 
discussed below, which refer only to the nontrivial case of inverted
hierarchy.

\begin{figure}[t]
\centering
\vspace*{-8mm}
\hspace*{15mm}
\epsfig{figure=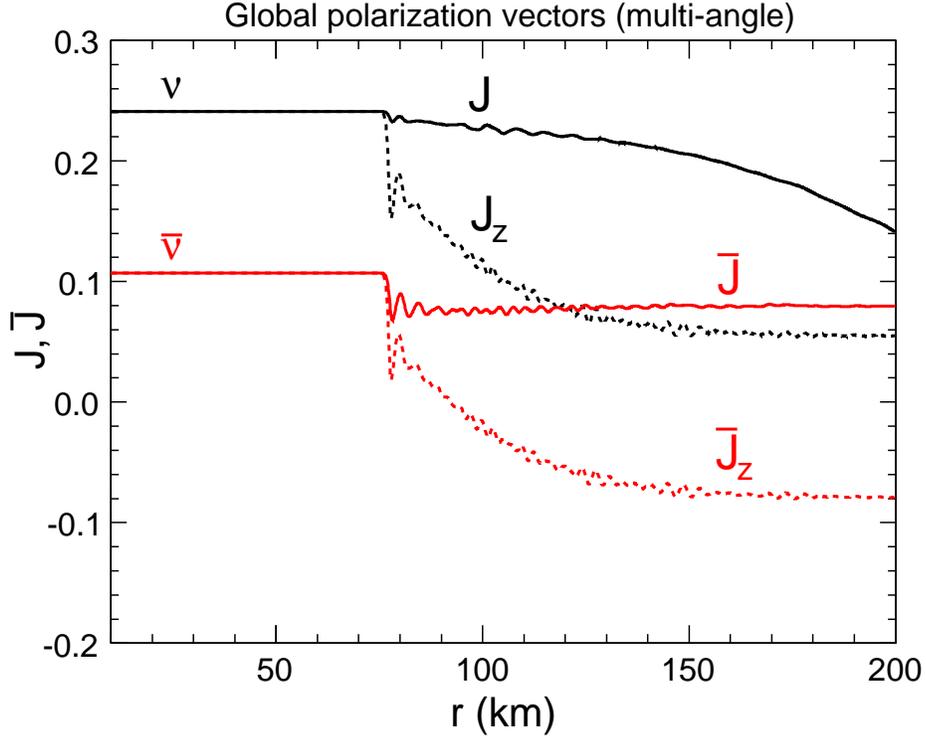,width =0.95\columnwidth}
\vspace*{-9.5mm}
\caption{\label{fig6} Multi-angle simulation in inverted hierarchy: Modulus and $z$-component of the global polarization vector
of neutrinos $(\bf J)$ and antineutrinos $(\overline{\bf J})$, as a function of radius.
The difference 
$D_z=J_z-\overline{J}_z$ (not shown) remains strictly constant in $r$.
}
\end{figure}

Figure~\ref{fig6} is the multi-angle analogue of Fig.~\ref{fig3}. By
comparing the two figures, it appears that 
bipolar oscillations of $\bf J$ and $\overline{\bf J}$ are largely 
suppressed in the multi-angle case, only
the first nutation being clearly visible. Moreover, such nutation
starts somewhat later ($r\simeq 76$~km) as compared with the single-angle
case ($r\simeq 68$~km). These features 
can be understood in terms of the different self-interaction effects experienced
along different trajectories. In multi-angle simulations,
neutrino-neutrino angles can be larger than the (single-angle) average one,
leading to somewhat stronger self-interaction effects, 
which keep the system in synchronized mode for a slightly longer time,
and thus delay the first nutation. Along different trajectories,
the subsequent bipolar oscillations have also somewhat different amplitudes and
phases, which tend to cancel out in the global polarization vectors.
For a similar reason
(relatively stronger self-interaction effects, as compared to single-angle), 
in Fig.~\ref{fig6} there is  a slightly more pronounced loss (depolarization) of $\overline J$
at the start of the first bipolar oscillation, and at the same time 
a prolonged ``coherence'' of $\bf {J}$ (slower decrease of $J$), 
with respect to Fig.~\ref{fig3}.
However, just as in the single-angle case, it 
turns out that $\overline{J}_z$ gets finally reversed, while the difference
$D_z=J_z-\overline{J}_z$ is exactly conserved. The reversal becomes more
evident by looking at specific 
(anti)neutrino energies.

\begin{figure}[t]
\centering
\vspace*{-3mm}
\hspace*{18mm}
\epsfig{figure=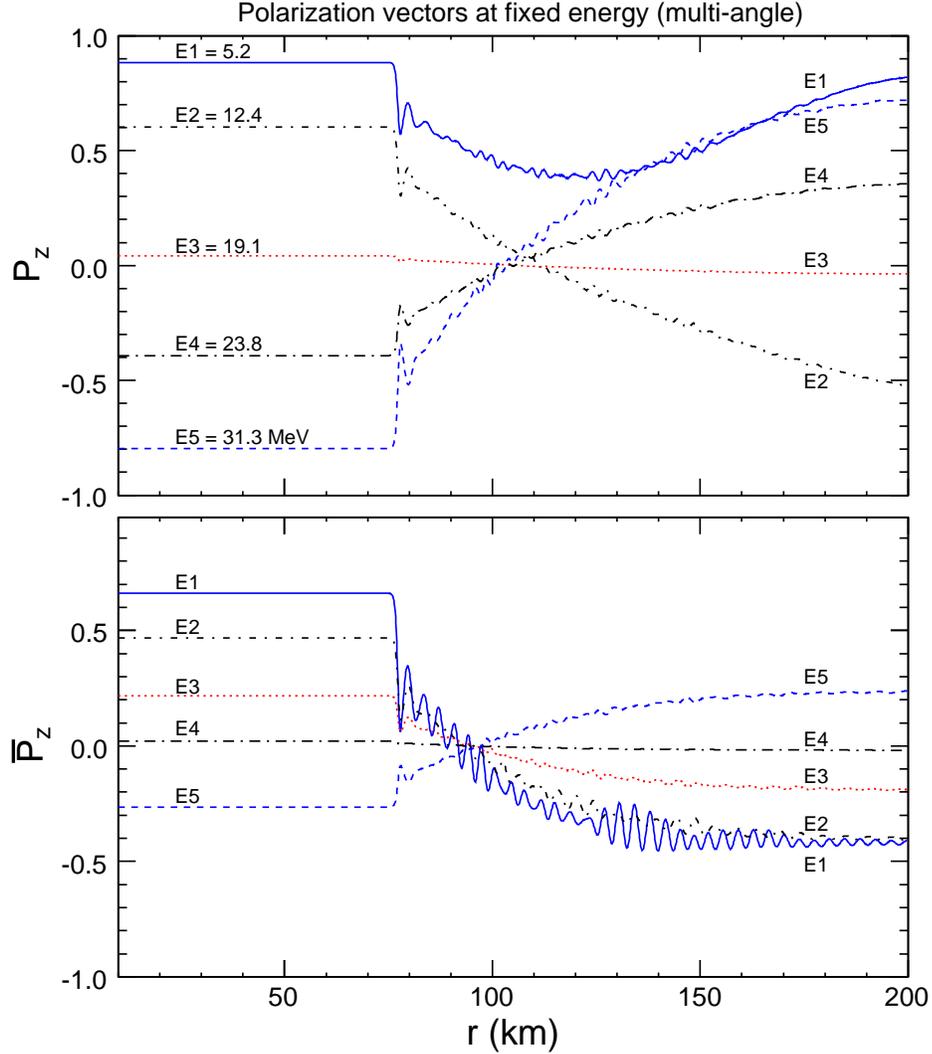,width =1.05\columnwidth}
\vspace*{-6mm}
\caption{Multi-angle simulation in inverted hierarchy: $z$-component of the polarization vector
of neutrinos ($P_z$, upper panel)  and antineutrinos ($\overline P_z$, lower panel)
as a function of radius, for five representative values of the energy.
\label{fig7}}
\end{figure}

Figure~\ref{fig7} shows the behavior of the $z$-component of angle-averaged
polarization vectors [Eq.~(\ref{angleave})] for neutrinos ($P_z$, upper panel)
and antineutrinos ($\overline{P}_z$, lower panel) in our multi-angle simulation,
at fixed energies. 
The behavior is qualitatively similar
to a ``smeared version'' of the single-angle curves in Fig.~\ref{fig4}, with
polarization vectors reversing (or not) their $z$-components as expected.  
The small depth (or absence) of nutations makes it more evident that
the polarization reversal (i.e., the spectral split) starts to dominate over 
the bipolar mode around the expected radius $r\simeq 100$~km.

\begin{figure}[t]
\centering
\vspace*{-4mm}
\hspace*{12mm}
\epsfig{figure=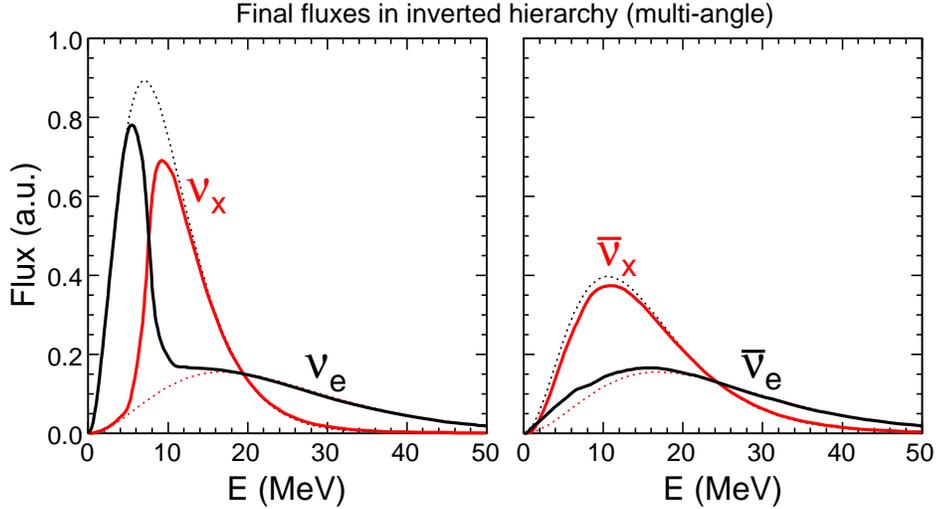,width =1.00\columnwidth}
\vspace*{-10mm}
\caption{Multi-angle simulation in inverted hierarchy: Final 
fluxes (at $r=200$~km, in arbitrary units) 
for different neutrino species as a function of energy. Initial fluxes are shown
as dotted lines to guide the eye.
\label{fig8}}
\end{figure}

Figure~\ref{fig8} shows the final ($r=200$~km) $\nu$ and $\overline\nu$
fluxes as a function of energy. The neutrino spectral swap at $E>E_c\simeq7$~MeV is rather
evident in the left panel, although it is less sharp  with respect to
the single-angle case in Fig.~\ref{fig5}. In the right panel of Fig.~\ref{fig8}, the
minor feature associated to the ``antineutrino spectral split'' 
 is largely smeared out (see the same panel in Fig.~\ref{fig5}), and 
survives as a small excess of $\overline\nu_e$ at low energy.

The spectra in Figure~\ref{fig8} are largely independent from the
specific mixing value chosen for the simulations ($\sin^2\theta_{13}=10^{-4}$), as far 
as $\theta_{13}>0$ (as we have also checked numerically). Variations
of $\sin^2\theta_{13}$ only lead to logarithmic variations
in the (unobservable) synchronized-bipolar transition radius, 
and in the depth of bipolar oscillations \cite{Hannestad:2006nj,Duan:2007mv}, which are anyway smeared out
in multi-angle simulations, as we have just seen. 
Therefore,
the spectra in Figure~\ref{fig8} may be taken as rather general
``initial conditions''  for possible later (ordinary or stochastic) matter 
effects, occurring when $\omega\sim \lambda(r)$ at $r\gg 200$~km. 
These later, ordinary matter effects are instead strongly dependent
on $\theta_{13}$, and vanish for, say, $\sin^2\theta_{13}\lesssim 10^{-5}$ 
(see, e.g., \cite{Foglish}).
If $\theta_{13}$ is indeed that small (but nonzero), neutrino self-interaction 
effects could be the only source of flavor transformations in (anti)neutrino spectra.

In conclusion, for $0<\sin^2\theta_{13}\lesssim 10^{-5}$,
the observable spectra at the SN exit would be similar to those
in Fig.~\ref{fig1} for the normal hierarchy case (no significant flavor
transformations of any kind), and to those in Fig.~\ref{fig8} for the
inverted hierarchy case (large self-interaction effects). For 
$\sin^2\theta_{13}\gtrsim 10^{-5}$, the same spectra should be
taken as ``initial conditions'' for the calculation of subsequent MSW effects.
Once more, we remark that the decoupling of self-interaction and MSW effects is a characteristic
of our adopted SN model, inspired by shock-wave simulations \cite{Foglish}.
The phenomenology becomes more complicated in alternative models with shallow
matter profiles,  when both effects can occur in the same region, as in the
simulations performed in \cite{Duan:2006an,Duan:2007bt}.


\section{Conclusions and prospects}

Neutrino-neutrino interactions in the high-density region near a supernova
core have been recently recognized to produce surprising collective effects.
Motivated by these developments, we have investigated
neutrino flavor transformations in a SN model, where the main
self-interaction effects (synchronization, bipolar 
oscillations, and spectral split) develop well before possible MSW effects.

The neutrino-neutrino interaction strength depends on the intersection angle
of their trajectories. Averaging this (variable) angle along a single, representative 
radial trajectory leads to the so-called single-angle approximation, which
allows both elegant analytical insights \cite{Hannestad:2006nj,Duan:2007mv} and easier numerical calculations. 
However, removal of this approximation
is needed (through multi-angle simulations) in order to validate the analytical insights,
and to show that the main effects are not spoiled by kinematical decoherence.
Moreover, many self-interaction effects have actually been first seen numerically
and then interpreted analytically {\em a posteriori}. 

We have thus performed numerical simulations in both single- and multi-angle
cases, using continuous energy spectra with significant $\nu$-$\overline \nu$
and $\nu_e$-$\nu_x$ asymmetry. The single-angle results can be understood
analytically to a large extent, and their main observable effect---in the
nontrivial case of inverted hierarchy---is the swap of energy spectra
above a critical energy dictated by lepton number conservation \cite{RaffSmirn}.
In multi-angle simulations, we find that the ``fine structure'' of
self-interaction effects (e.g., bipolar oscillations) is smeared out,
but the spectral swap remains a robust, observable feature. In this sense,
trajectory averaging does not play a crucial role. This is
the main result of our work.

The swapping of the $\overline{\nu}_e$ and
$\overline{\nu}_\mu$ (as well as of the $\nu_e$ and $\nu_\mu$) fluxes could have 
an impact on r-process nucleosynthesis~\cite{Pantaleone:1994ns,%
Qian:1994wh,Sigl:1994hc,Qian:1993dg}, on the energy transfer to the
stalling shock wave~\cite{Fuller1992}, and on the possibility to observe
shock-wave propagation effects in neutrinos.
It is also worth studying possible self-interaction
effects in the phenomenology of SN~1987A neutrino 
events~\cite{Mirizzi:2005tg,Lunardini,Yuksel:2007mn} 
and of the diffuse supernova neutrino background spectrum~\cite{Malek:2002ns,Foglirelic}.
Further analytical and numerical developments may require to solve the
neutrino evolution equations in the
complete $3 \nu$ flavor scenario, where new effects associated to the ``solar''
$\delta m^2$ can occur; a recent example has been worked out in \cite{DuanLast}. 
Perturbations
of the (cylindrically symmetric) bulb model for neutrino emission might
also be considered in more advanced simulations.
Finally, there is a continuing interest in more formal aspects
of the mean-field approach to neutrino self-interaction effects \cite{Friedland:2003eh},
which has been implicitly assumed in most of the related literature (including this work); 
see \cite{Balantekin:2006tg}
for a recent discussion of its validity and inherent approximations.

In conclusion, twenty years after the SN1987A, the understanding
of SN $\nu$ flavor transformations is still in progress,
and surprising  self-interaction effects
are emerging as possible dominant phenomena. 
These effects are changing the current paradigm of SN neutrino physics, 
and demand further analytical and numerical investigations,
as well as new experimental inputs and guidance---should Nature be so kind to make a
galactic supernova explode.

\section*{Acknowledgments}
 
This work is supported in part by 
the Italian ``Istituto Nazionale di Fisica Nucleare'' (INFN) and by the ``Ministero dell'Universit\`a e 
della Ricerca'' (MiUR) through the ``Astroparticle Physics'' research project.
In Germany, the work of A.M.\  is supported by an Alexander~von~Humboldt fellowship grant.

We thank F.~Iavernaro and F.\ Mazzia for invaluable help in implementing their GAMD software
\cite{GAMD}.
A.M. thanks G.G.~Raffelt for illuminating discussions. 
We acknowledge D.~Montanino and
G.G.~Raffelt for reading the manuscript and for useful remarks, and
B.~Dasgupta, H.~Duan, G.M.~Fuller, and C.~Lunardini for communications.


\begin{figure}[t]
\centering
\vspace*{-7mm}
\hspace*{-15mm}
\epsfig{figure=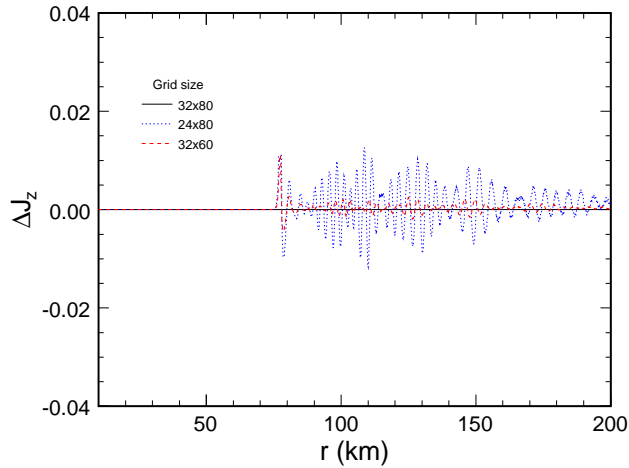,width =0.62\columnwidth}
\vspace*{-5mm}
\caption{Absolute difference between numerical evaluations of
$J_z(r)$ in multi-angle simulations, using various (energy)$\times$(angle)
grid sizes: $32\times 80$ (baseline), $32\times 60$ (dashed), and $24\times 80$ (dotted).
\label{fig9}}
\end{figure}

\appendix
\section*{Appendix}

We discuss here a few technical aspects of multi-angle numerical
simulations, which are much more challenging than single-angle ones. Equations~(\ref{bloch3},\ref{bloch4}),
after the proper transformation $\vartheta\to\vartheta_0$,
provide a set of $6\times N_E\times N_{\vartheta_0}$ ordinary differential equations (ODE)
in $r$ (in the real domain), where $N_E$ and $N_{\vartheta_0}$ are the number of points sampling the
(anti)neutrino energy $E$ and emission angle cosine $\cos\vartheta_0$, respectively.

This ODE set is stiff, and needs appropriate routines to be solved 
numerically. After a careful comparison of public routines, we have adopted the GAMD software \cite{GAMD}, 
implemented on a $N_E\times N_{\vartheta_0}=32\times 80$ grid. Denser sampling  
in $\cos\vartheta_0$ is required, since the polarization vectors generally vary
more rapidly in $\vartheta_0$ than in $E$. An exception would be provided by
MSW effects interfering with self-interaction ones \cite{Duan:2006an} (not our case), which
requires much denser sampling, especially in $E$, in order to track
the MSW resonant behavior. In our scenario with no MSW interference,
we obtain satisfactory numerical convergence 
with $N_E\times N_{\vartheta_0}=32\times 80$. Figure~\ref{fig9}
compares the last three steps in a trial sequence with increasingly denser
grids $N_E\times N_{\vartheta_0}$, showing that the final absolute error 
on the reference quantity $J_z$ can be safely estimated to be  $<10^{-2}$. 
  
The grid points are not chosen to be equally spaced, but are instead fixed by imposing
weighted Gaussian quadrature of the double integrals in the right-hand-side of
Eqs.~(\ref{bloch3},\ref{bloch4}). In other words, the $N_E\times N_{\vartheta_0}$
grid points not only sample the energy and angular evolution of the polarization
vectors in the ODE set, but are also used to perform the inner Gaussian integrations
at each evolutionary step. This ``trick'' saves a lot of computer time. Nevertheless,
a typical simulation over a $32\times 80$ grid takes $\sim 400$ hours
on our local computer facility (a cluster of four, 64-bit and 4Gb RAM processors
at 2.4 GHz, with Fortran~90 codes running on a Linux platform). We plan to use 
more powerful (remote) facilities in future works on the subject.

\section*{References} 

\end{document}